\documentclass[conference]{IEEEtran}
\IEEEoverridecommandlockouts

\usepackage{cite}
\usepackage{amsmath,amssymb,amsfonts}
\usepackage{algorithmic}
\usepackage{graphicx}
\usepackage{textcomp}
\usepackage{xcolor}
\usepackage{balance}
\usepackage{multirow}
\usepackage{booktabs}
\newcommand{\guille}[1]{``#1''}
\def\BibTeX{{\rm B\kern-.05em{\sc i\kern-.025em b}\kern-.08em
    T\kern-.1667em\lower.7ex\hbox{E}\kern-.125emX}}
\begin{document}

\title{HermiCache: Enclave-Aware Cache Replacement for Trusted Execution Environments\thanks{This work was carried out within the SCAMA project (ANR-23-CE39-0011),
funded by the French National Research Agency (ANR).}
}

\author{\IEEEauthorblockN{Oussama Elmnaouri\IEEEauthorrefmark{1}, Pascal Cotret\IEEEauthorrefmark{1}, Vianney Lapôtre\IEEEauthorrefmark{2}, Loïc Lagadec\IEEEauthorrefmark{1}}
\IEEEauthorblockA{\IEEEauthorrefmark{1} Lab-STICC, UMR CNRS 6285, ENSTA (29806 Brest Cedex 9, France)\\firstname.lastname@ensta.fr}
\IEEEauthorblockA{\IEEEauthorrefmark{2} Lab-STICC, UMR CNRS 6285, Université Bretagne-Sud (56100 Lorient, France)\\vianney.lapotre@univ-ubs.fr}
}

\maketitle

\begin{abstract}
Trusted Execution Environments (TEEs) protect enclave memory from untrusted software but remain vulnerable to cache-based side-channel attacks due to shared microarchitectural resources. Existing countermeasures use techniques such as cache partitioning or randomization: these solutions are not ideal if a designer wants fine-grained configurations and a deterministic protection. In this paper, we introduce HermiCache which is an answer to these requirements. HermiCache is designed for RISC-V cores and has been implemented in the OpenHwGroup CVA6 core with a Keystone TEE for the software layer. The solution has an area overhead of 6\% on the processor core.
\end{abstract}


\section{Introduction}
Trusted Execution Environments (TEEs) ensure that sensitive code and data are processed in enclaves which isolate it from untrusted software. Processor architectures usually have one or more TEEs available: Intel SGX \cite{2016_Costan_Intel-SGX}, ARM TrustZone \cite{2016_Ngabonziza_ARM-TrustZone}, and AMD SEV \cite{2020_AMD_SEV-snp}. For the open-source RISC-V architecture, Keystone \cite{2020_Lee_Keystone} and Penglai \cite{2021_Feng_Penglai} are two possible solutions. 

TEEs do not only rely on software components, it also implies that hardware mechanisms which microarchitecture can leak information. When using shared caches, timing side-channel attacks such as Evict+Time or Prime+Probe \cite{2018_Ge_Survey-timing-attacks} can be used to retrieve information from the victim. Existing defenses can be software-based or hardware-based, but both engender important tradeoffs: techniques such as constant-time programming are difficult to use with complex applications while hardware partitioning or randomization implies a non-optimized capacity and additional logic to manage the cache replacement policy.

This paper introduces HermiCache, a hardware-enforced isolation for TEEs resistant to cache-based side-channel attacks. The contributions are as follows:

\begin{itemize}
    \item Ownership-based replacement mechanism.
    \item RTL implementation in a RISC-V CVA6 core with a Keystone TEE integration.
    \item Quantitative evaluation through FPGA synthesis and gem5 simulations.
\end{itemize}

\section{HermiCache architecture}

\subsection{Threat model}
HermiCache follows the Keystone threat model and extends it to cache-based side-channel attacks. The processor, the cache hierarchy and the Keystone Security Monitor (SM) are trusted while other software components are not (operating system or REE applications, for instance). The attacker is allowed to perform cache attacks (such as Prime+Probe, Evict+Time and Flush+Reload) and execute untrusted code. HermiCache does not protect against all microarchitectural side-channels (i.e. TLB, branch predictor) and does not take into account physical attacks.

\subsection{Design}

\begin{figure}[htbp]
    \centering
    \includegraphics[width=0.81\linewidth]{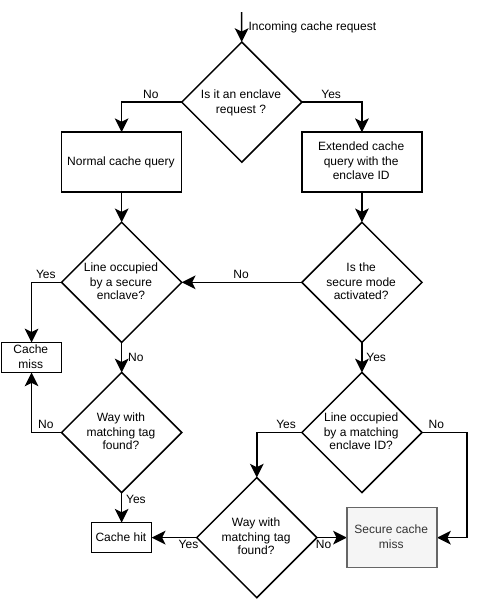}
    \caption{HermiCache ownership-aware lookup and replacement policy.}
    \label{fig:ReplacementPolicy}
\end{figure}

HermiCache extends a set-associative cache with ownership metadata for each cache line: a 1-bit security flag (\guille{is the line protected by HermiCache?}) and an Enclave IDentifier (\guille{which enclave is linked to the current line?}). The cache replacement policy is described in Figure \ref{fig:ReplacementPolicy}. For a secure cache miss (gray block), it follows this priority order: 1) select an invalid entry; 2) select a line of the current enclave; 3) select a non-protected line; 4) if no eligible victim exists, bypass allocation at this cache level.

\begin{figure}[htbp]
    \centering
    \includegraphics[width=\linewidth]{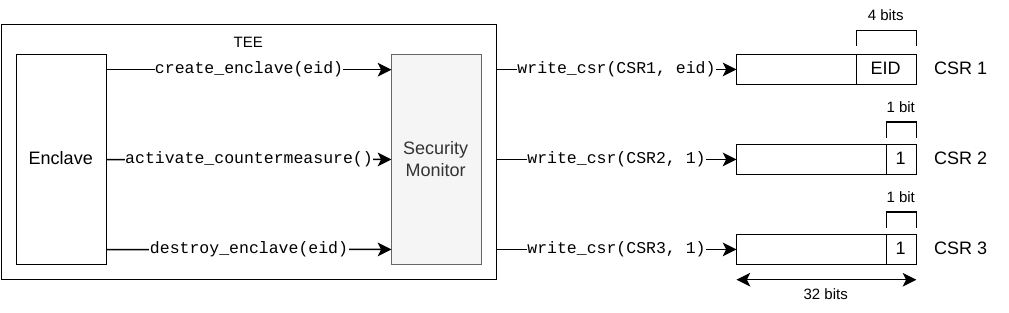}
    \caption{Security Monitor configuration of HermiCache ownership state through trusted CSRs.}
    \label{fig:hardware-software-microarchitecture}
\end{figure}

Regarding the integration within a Keystone TEE environment, HermiCache extends the API with functions allowing an enclave to request an explicit cache protection (see Figure \ref{fig:hardware-software-microarchitecture}). The Keystone SM assigns an EID and configures the security flag, both information are then stored in trusted CSR registers added in the CVA6 core. By default, HermiCache is configured with a 4-bit EID allowing up to 15 protected identifiers (+1 dedicated to non-secure execution).

\section{Evaluation}

HermiCache is implemented for a CVA6-based Keystone SoC on a Digilent Genesys2 board (XC7K325T FPGA) with Vivado 2018.3. The SoC includes some peripherals as well as a 16KB 4-way L1 instruction cache, a 32KB 8-way L1 data cache and a DDR3 memory. Implementation results are given in Table \ref{tab:hw_overhead}.

\begin{table}[htbp]
\centering
\caption{Post-implementation resource, timing, and power overhead of HermiCache.}
\label{tab:hw_overhead}
\resizebox{\linewidth}{!}{
\begin{tabular}{lcccc}
\toprule
\textbf{Metric} & & \textbf{Baseline} & \textbf{HermiCache} & \textbf{Overhead} \\
\midrule
\multirow{2}{*}{LUTs} & Full SoC   & 71,326  & 72,236 & \(+1.27\)\% \\
& Core only & 46,453 & 47,358 & \(+1.94\)\% \\\midrule
\multirow{2}{*}{FFs} &  Full SoC   & 45,573  & 46,985 & \(+3.1\)\% \\
 & Core only   &  23,375 & 24,791 & \(+6.05\)\% \\\midrule
\multirow{2}{*}{BRAMs} &  Full SoC & 53  & 53 & \(0.0\)\% \\
 & Core only &  36 & 36 & \(0.0\)\% \\\midrule
 \multicolumn{2}{c}{Total Power (W)} & 6.75 & 6.887 & \(+2.02\)\%\\
 \midrule
 \multicolumn{2}{c}{Worst Timing Slack (ns)} & 0.177 & 0.177 & \(0.0\)\%\\
\bottomrule
\end{tabular}
}
\end{table}

HermiCache increases LUT usage by 1.27\% and flip-flop usage by 3.1\%, with no additional BRAMs as the approach only add metadata and control logic. The latency on a cache hit or a cache miss is unchanged compared to a baseline cache. In a same-set contention experiment, attacker-induced eviction of a protected victim line succeeded in 100\% of trials with LRU replacement and 58\% with random replacement, but 0\% with HermiCache.

\section{Related work}

\begin{figure}[htbp]
    \centering
    \includegraphics[width=0.9\linewidth]{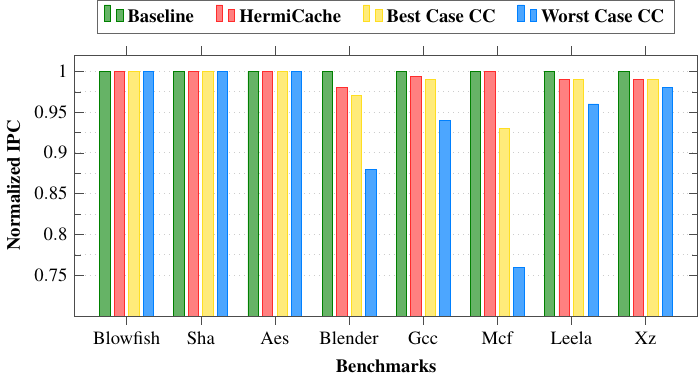}
    \caption{Configuration-aligned normalized IPC comparison between HermiCache and Composable Cachelets (CC) with several benchmarks.}
    \label{fig:Hermi_Vs_CC}
\end{figure}

There are several cache isolation mechanisms for TEEs \cite{2020_Dessouky_Hybcache,2022_Daniel_Composable_Cachelets,2025_Yin_VeriCache}. HermiCache is compared with Composable Cachelets (CC) \cite{2022_Daniel_Composable_Cachelets} as it is a recent TEE-oriented cache-isolation mechanism based on fine-grained cache-capacity allocation. Figure \ref{fig:Hermi_Vs_CC} shows that HermiCache is competitive compared to the best and worst CC configurations when regarding the Instructions Per Cycle (IPC) rate.

\section{Conclusion}
This paper presented HermiCache which is a cache protection mechanism for TEEs. It has low area and IPC overheads. In future works, HermiCache will be extended with broader attack coverage and scalability analysis as well as more benchmarks in order to study the efficiency of HermiCache against several workloads.

\bibliographystyle{IEEEtran}
\bibliography{biblio}

\end{document}